\documentclass{article}

 \usepackage[dblblindworkshop, final]{neurips_2025}
\workshoptitle{The First Workshop on Generative and Protective AI for Content Creation}

\usepackage[utf8]{inputenc} % allow utf-8 input
\usepackage[T1]{fontenc}    % use 8-bit T1 fonts
\usepackage{hyperref}       % hyperlinks
\usepackage{url}            % simple URL typesetting
\usepackage{booktabs}       % professional-quality tables
\usepackage{amsfonts}       % blackboard math symbols
\usepackage{nicefrac}       % compact symbols for 1/2, etc.
\usepackage{microtype}      % microtypography
\usepackage{xcolor}         % colors
\usepackage{graphicx}
\usepackage{subcaption}
\usepackage{longtable}

\title{Decomate: Leveraging Generative Models for Co-Creative SVG Animation}

\author{%
  Jihyeon Park$^{*}$ \\
  MODULABS\\
  \texttt{milhaud1201@gmail.com} \\
  \And
  Jiyoon Myung$^{*}$ \\
  MODULABS\\
  \texttt{jiyoon0424@gmail.com} \\
  \And
  Seone Shin \\
  MODULABS \\
  \texttt{tttlstjsdliiii@gmail.com} \\ 
  \And
  Jungki Son \\
  MODULABS \\
  \texttt{aeolian83@gmail.com} \\
  \And
  Joohyung Han \\
  MODULABS \\
  \texttt{ddang8jh@gmail.com} \\
}

\begin{document}

\maketitle

\begingroup
\renewcommand\thefootnote{*}
\footnotetext{Equal contribution.}
\endgroup

\begin{abstract}
Designers often encounter friction when animating static SVG graphics, especially when the visual structure does not match the desired level of motion detail. Existing tools typically depend on predefined groupings or require technical expertise, which limits designers’ ability to experiment and iterate independently. We present Decomate, a system that enables intuitive SVG animation through natural language. Decomate leverages a multimodal large language model to restructure raw SVGs into semantically meaningful, animation-ready components. Designers can then specify motions for each component via text prompts, after which the system generates corresponding HTML/CSS/JS animations. By supporting iterative refinement through natural language interaction, Decomate integrates generative AI into creative workflows, allowing animation outcomes to be directly shaped by user intent.
\end{abstract}

\section{Introduction}
Animating static SVGs remains a technically demanding task for UI/UX designers. Despite modern design tools, designers must still navigate multiple platforms, manually edit code, or rely on developers to implement motion. These workflows often lack flexibility, fine-grained control, and iterative creative feedback.

Recent work explores how generative models can support creative content creation in design workflows. For example, Kolthoff et al. investigate zero-shot prompting methods for generating high-fidelity GUI prototypes from natural language descriptions~\cite{kolthoff2024zeroshot}, and Xing et al. present datasets that enhance LLMs’ capabilities for SVG understanding and generation~\cite{xing2024empowering}. In animation specifically, Keyframer enables designers to generate SVG animations via natural language prompts, but it assumes well-structured assets with explicit class labels~\cite{keyframer}. These approaches illustrate promising directions yet fall short of addressing the gap between unstructured design assets and expressive, user-driven motion.

We introduce \textbf{Decomate}, an interactive system that enables designers to animate SVGs by combining LLM-powered semantic decomposition with language-based animation specification. The system restructures unorganized SVGs into animation-relevant components, suggests motion prompts per group, and allows users to refine both structure and motion via natural language. The output is production-ready HTML/CSS/JS, supporting a co-creative, low-friction animation workflow.
Decomate reimagines how designers and intelligent systems collaborate—shifting from static structure to expressive animation through shared interpretation and iterative feedback.

\section{Motivation}

To better understand designers’ pain points, we conducted interviews with 11 professional product and UX designers across a range of experience levels (Table~\ref{tab:participants}). Several consistent themes emerged.

First, animation is frequently deprioritized due to time constraints and rigid team schedules. Even designers who considered motion important for clarity and user engagement reported limited opportunities to explore or integrate it meaningfully.

Second, fine-tuning animations—such as adjusting timing, easing, or expressive behaviors—was described as tedious and overly dependent on developer implementation. Many participants lacked direct control over motion refinement, leading to a disconnect between their creative vision and the final result.

Third, while all participants expressed interest in AI-assisted animation, none had experience using such tools. This was largely due to the absence of approachable, design-oriented solutions and a lack of familiarity or confidence in formulating effective prompts.

These findings underscore the need for a system that minimizes technical barriers while maximizing expressive control. Decomate responds to this need by enabling natural language-based animation authoring and iterative refinement, empowering designers to explore motion as a reusable, flexible design material.

\begin{figure*}[t]
    \centering
    \includegraphics[width=\linewidth]{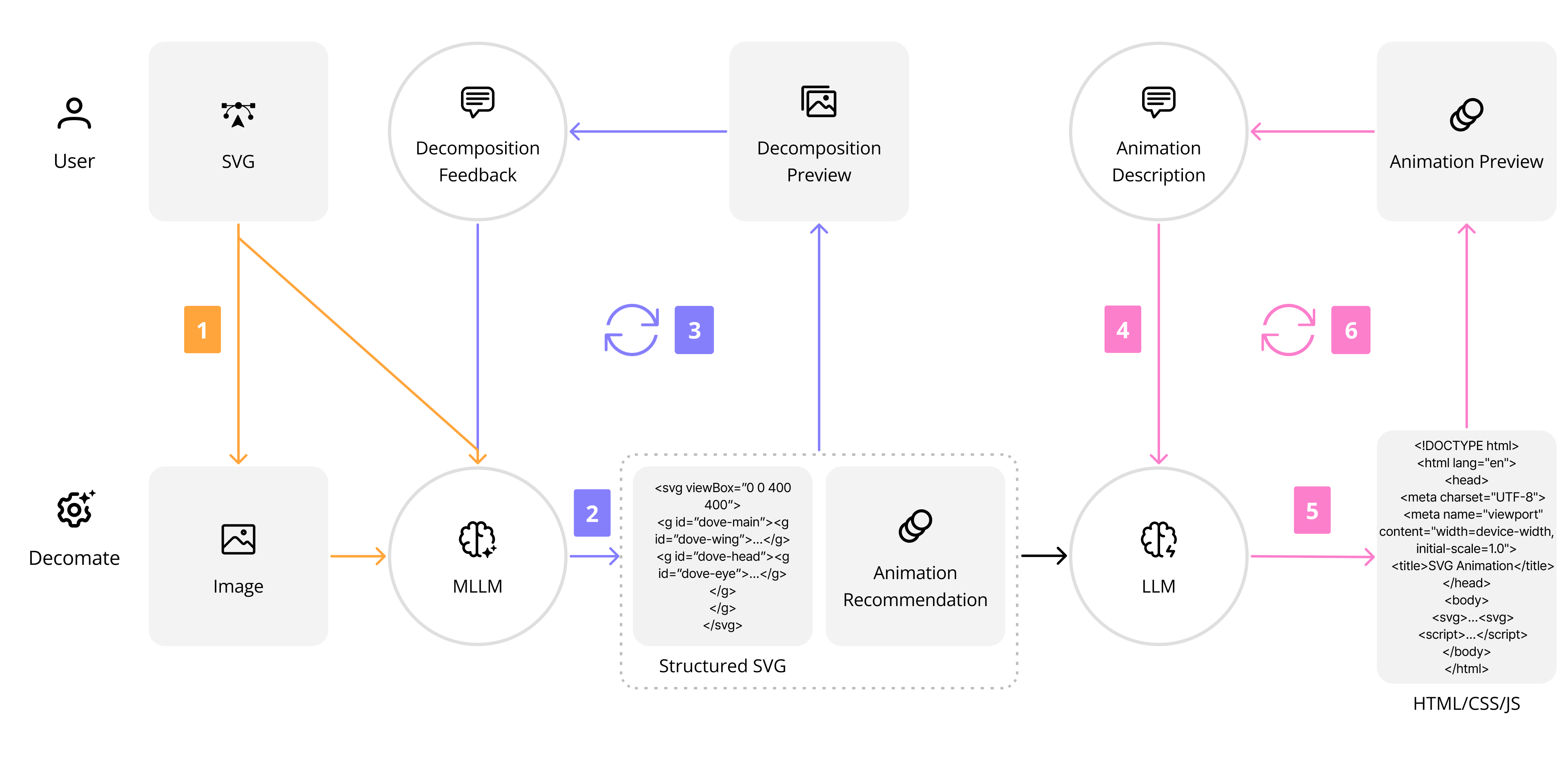}
    \caption{
    Overview of the Decomate pipeline. Users provide an SVG file or code, which is semantically grouped by a multimodal LLM. Groupings can be refined via natural language. After selecting or describing desired motions, an LLM generates corresponding HTML/CSS/JS code and preview, enabling iterative co-creation.
    }
    \label{fig:decomate_flow}
\end{figure*}

\section{Methodology}

\begin{figure*}[]
    \centering
    \includegraphics[width=\linewidth]{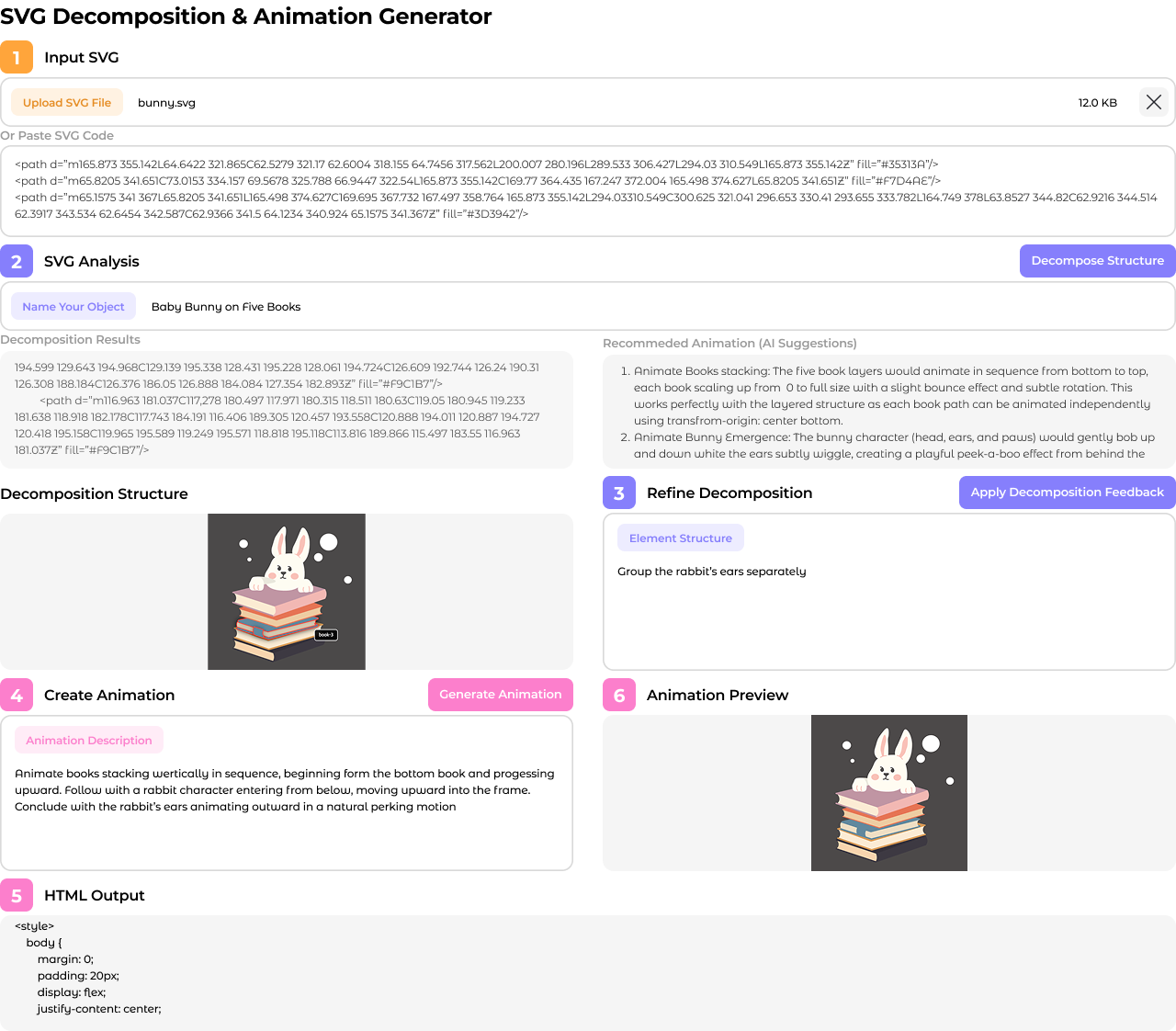}
    \caption{
    \textbf{Decomate user interface.} The system enables designers to animate static SVGs through an interactive language-driven pipeline:
    \textbf{(1)} Users upload or paste SVG code. 
    \textbf{(2)} A multimodal LLM analyzes the SVG and suggests semantic groupings and animation ideas. 
    \textbf{(3)} Users can refine decomposed elements using natural language. 
    \textbf{(4)} Animation behavior is described through prompts.
    \textbf{(5)} The system generates HTML/CSS/JS code with animation.
    \textbf{(6)} A live preview allows users to inspect results and iteratively refine motion.
    }
    \label{fig:decomate_ui}
\end{figure*}

Decomate is an interactive pipeline for animating static SVGs through semantic structuring and natural language interaction. It integrates visual analysis, multimodal large language model (LLM) inference, and user feedback into a co-creative loop. Figure~\ref{fig:decomate_flow} provides an overview of the full pipeline. Figure~\ref{fig:decomate_ui} illustrates the user interface corresponding to each step in the pipeline. A live demo, along with sample animation GIFs, is available online\footnote{\url{https://huggingface.co/spaces/PrompTartLAB/Decomate}}.

\paragraph{SVG Input}  
Users begin by uploading an SVG file or pasting raw SVG code. The input may range from well-structured to entirely flat graphics, with or without class names or groupings. Decomate is designed to handle this variability and requires no prior asset preparation. To assist semantic grouping, users specify a high-level object name (e.g., “dog”), which guides the system in decomposing the graphic into meaningful parts (e.g., ears, nose, mouth, legs).

\paragraph{Semantic Decomposition and Animation Suggestions}  
A multimodal large language model (MLLM; Claude Sonnet 4) analyzes both the SVG code and its rendered image to identify animation-relevant components. It restructures the SVG into semantically meaningful groups, prioritizing visual interpretation over existing syntactic hierarchies. For each group, the model also generates preliminary animation suggestions in natural language.

\paragraph{Preview and Structural Refinement}  
Users review the proposed groupings in a visual preview. If the structure does not align with their creative intent, they can provide natural language feedback (e.g., “split the left and right feet”). The system uses this feedback to iteratively revise the semantic grouping, ensuring alignment between designer intent and structural representation.

\paragraph{Prompt-Based Animation Authoring}  
Using the system’s animation suggestions as a baseline, users specify desired motion behaviors through free-form natural language prompts (e.g., “make the wings flap slowly with elastic easing”). This enables expressive and intuitive animation authoring without requiring technical syntax.

\paragraph{Code Generation and Rendering}  
A text-based language model (LLM; Claude Sonnet 4) translates the final group structure and animation prompts into production-ready HTML and CSS. The system outputs a fully animated SVG along with the corresponding HTML/CSS/JS code, allowing users to create animations without dealing with complex technical details.

\paragraph{Interactive Preview and Iteration}  
The animation is rendered in a live preview pane. Users can inspect timing, easing, or spatial behavior, and refine the animation further through additional prompts (e.g., “increase the bounce on landing”). Each user revision triggers regeneration of the animation code and preview, enabling expressive iteration through natural interaction.

\section{User Study}

We conducted a formative study with six participants to evaluate Decomate's usability and creative support. 
Participants animated an SVG using the system and shared qualitative feedback on intent alignment, SVG decomposition, and visual naturalness (see Appendix~\ref{tab:user_feedback}).

\paragraph{Positive Feedback}
Participants found the semantic decomposition generally effective, especially when dealing with flat or unstructured SVGs. They appreciated that minimal effort was needed to generate animations and that the system supported interactive refinement of groupings via natural language. This co-creative loop made it easier to match the animation structure to their creative intent without manual editing.

\paragraph{Challenges and Improvement Directions}
The most common challenge was specifying animations via natural language. While the system’s suggestions helped, participants were often unsure how to phrase prompts. In addition, several participants mentioned that it was challenging to express or manipulate fine-grained animation details—such as subtle motion intensity or layer-level timing—to achieve their intended effects. 

To address this, we aim to provide prompt examples and templates to reduce the learning curve for natural language prompting. Furthermore, we plan to implement an LLM-driven suggestion that predicts user intent from partial inputs and recommends complete prompt candidates. Finally, we plan to incorporate structured animation controls (e.g., sliders and toggles) that allow for precise tuning of behaviors such as timing, easing, or direction.

\section{Conclusion}

We presented Decomate, a system that enables designers to animate static SVG graphics through an interactive, language-driven workflow. By combining semantic decomposition, prompt-based animation suggestion, and iterative refinement through natural language, Decomate allows designers to focus on what they want to express rather than how to technically implement it. The system reframes animation as a co-creative process in which language and computation jointly shape visual motion, emphasizing human intent and authorship while leveraging the adaptability of large language models. Based on insights from our user study, we plan to enhance Decomate’s ability to reflect users’ creative intent more accurately, introduce finer-grained animation controls and a wider range of motion styles, and expand prompt guidance to reduce the learning curve and support more expressive outcomes.

\begin{ack}
This research was supported by Brian Impact Foundation, a non-profit organization dedicated to the advancement of science and technology for all.
\end{ack}

\bibliographystyle{plain} 
\bibliography{decomate}

\newpage
\appendix

\section{Formative Interview Participants}

We conducted formative interviews with 11 professional product designers to understand their workflows and challenges around motion design in UI. The table below summarizes their backgrounds.

\begin{table}[h]
\centering
\begin{tabular}{c|l|l|l|l}
\hline
\textbf{ID} & \textbf{Job Title} & \textbf{Years of Experience} & \textbf{Animation Tools} & \textbf{Gender} \\
\hline
P1 & Product Designer & 5–10 years & Code (basic) & Female \\
P2 & Product Designer & 3–5 years & Aninix, Figma, Lottie & Female \\
P3 & Product Designer & 5–10 years & Figma & Female \\
P4 & Product Designer & 1–3 years & After Effects, Figma & Female \\
P5 & Product Designer & 5–10 years & After Effects, Figma, Lottie  & Male \\
P6 & Product Designer & 3–5 years & Figma & Female \\
P7 & UX Designer & 5–10 years & Figma, Lottie  & Female \\
P8 & UX Designer & 3–5 years & Figma, Lottie & Female \\
P9 & UX Designer & 5–10 years & Figma, Lottie & Male \\
P10 & UX Designer & 1–3 years & After Effects, Figma  & Female \\
P11 & UX Designer & 3–5 years & Figma, Lottie & Female \\
\hline
\end{tabular}
\caption{Formative interview participants: job roles, experience, and tool usage.}
\label{tab:participants}
\end{table}

\begin{longtable}{p{0.5cm} p{13cm}}
\caption{Qualitative feedback from user study participants (translated excerpts).}
\label{tab:user_feedback} \\

\hline
\textbf{ID} & \textbf{Responses} \\
\hline
\endfirsthead

% 빈 head (헤더 반복 없음)
\endhead

P1 & 
\textbf{Alignment with Intent:} Wanted wing movement for “flying bird,” but unrelated parts moved or detached. The attempt to simulate vertical motion was visible, though results were misaligned with intent.\\
& \textbf{SVG Structure:} Placement followed the file, but grouping was incomplete, leading to awkward motion.\\
& \textbf{Naturalness:} UX was difficult without prior experience; motion felt rough.\\
& \textbf{Visual Quality:} Despite awkwardness, the intent was clear (vertical movement). Could be useful if quality improves.\\
\hline

P2 & 
\textbf{Alignment with Intent:} Object decomposition was good, but animation behaved unexpectedly (e.g., whale’s eye detached).\\
& \textbf{SVG Structure:} Need clearer rules for what to separate (droplets) vs. keep together (eyes, arms).\\
& \textbf{Naturalness:} Explanation or presets could help users input prompts directly.\\
& \textbf{Visual Quality:} Good droplet effect; richer motions (rotation, scale, curve paths) would improve realism.\\
\hline

P3 & 
\textbf{Alignment with Intent:} The rocket’s upward motion matched the request, but its parts moved separately instead of cohesively.\\
& \textbf{SVG Structure:} Some static parts were incorrectly animated.\\
& \textbf{Naturalness:} Timing inconsistency—some parts too fast or abrupt transitions.\\
& \textbf{Visual Quality:} Physics-based easing was good; consistency could be improved.\\
\hline

P4 & 
\textbf{Alignment with Intent:} The output differed from the prompt.\\
& \textbf{SVG Structure:} Logical and animation-ready.\\
& \textbf{Naturalness:} Some awkward transitions.\\
& \textbf{Visual Quality:} Acceptable overall flow, but some incomplete parts.\\
\hline

P5 & 
\textbf{Alignment with Intent:} Easy to get SVG animation from text alone, but without AI suggestions, motion was awkward.\\
& \textbf{SVG Structure:} Grouping was fine in sample tests.\\
& \textbf{Naturalness:} Tried “whale fin moves left and right,” but motion was linear, not swinging; prompting had a learning curve.\\
& \textbf{Visual Quality:} Easy generation is strong; finer control desired. Suitable for small loop animations.\\
\hline

P6 & 
\textbf{Alignment with Intent:} Example used “whale” and “rocket”; roughly matched, but grouping issues hindered evaluation.\\
& \textbf{SVG Structure:} Whale body, eye, and features moved separately—should be grouped; bubbles should be distinct.\\
& \textbf{Naturalness:} Layout of “Decomposed Elements / Create Animation” panels was confusing.\\
& \textbf{Visual Quality:} Direct UI controls (curve, duration) would improve usability beyond text-only input.\\
\hline

\end{longtable}

\end{document}